\documentclass[prl,amsmath,amssymb,twocolumn, showpacs, superscriptaddress,10pt]{revtex4-1} 
\usepackage{amsmath}
\usepackage{hyperref}
\usepackage{graphicx}
\usepackage{amsfonts}
\usepackage{amsthm}
\usepackage{cases}
\usepackage{bm}
\usepackage[normalem]{ulem}

\newcommand{\be}{\begin{equation}}
\newcommand{\ee}{\end{equation}}
\newcommand{\bea}{\begin{eqnarray}}
\newcommand{\eea}{\end{eqnarray}}

\usepackage{color}
\definecolor{Blue}{rgb}{0.00, 0.00, 1.00}
\definecolor{Red}{rgb}{1.00, 0.00, 0.00}
\definecolor{Green}{rgb}{0.00, 0.70, 0.00}

\newcommand{\blue}{\color{Blue}}

\begin{document}

\title{A mean-field theory for heterogeneous random growth with redistribution}

\author{Maximilien Bernard}
\affiliation{Laboratoire de Physique de l'\'Ecole Normale Sup\'erieure, CNRS, ENS $\&$ PSL University, Sorbonne Universit\'e, Universit\'e Paris Cité, 75005 Paris, France}
\affiliation{LPTMS, CNRS, Univ. Paris-Sud, Universit\'e Paris-Saclay, 91405 Orsay, France}
\author{Jean-Philippe Bouchaud}
\affiliation{Capital Fund Management, 23 rue de l'Université, 75007 Paris, France \& Académie des Sciences, 75006 Paris, France}
\author{Pierre Le Doussal}
\affiliation{Laboratoire de Physique de l'\'Ecole Normale Sup\'erieure, CNRS, ENS $\&$ PSL University, Sorbonne Universit\'e, Universit\'e Paris Cité, 75005 Paris, France}

\date{\today}

\begin{abstract}
We study the competition between random multiplicative growth and redistribution/migration in the mean-field limit, when the number of sites is very large but finite. We find that for static random growth rates, migration should be strong enough to prevent localisation, i.e. extreme concentration on the fastest growing site. In the presence of an additional temporal noise in the growth rates, a third partially localised phase is predicted theoretically, using results from Derrida's Random Energy Model. Such temporal fluctuations mitigate concentration effects, but do not make them disappear. We discuss our results in the context of population growth and wealth inequalities.       
\end{abstract}

\maketitle

Models of random multiplicative growth with ``redistribution'' or ``migration'' naturally appear in a very large variety of contexts (population growth, ecology, immunology, genetics, economics and finance) and, in disguise, in statistical mechanics (directed polymers in random media, stochastic surface growth and the associated KPZ equation \cite{KPZ,HHZ}). Consider, for example, the dynamics of the populations of cities \cite{gabaix,barthelemy}. Climate, economic, or health factors may all fluctuate in time, leading to random multiplicative growth which is also coupled to migrations away from and into each city. Depending on the setting, one may either observe ``localisation'' (a.k.a ``condensation''), where most of the population lives in one or a handful of cities, or ``delocalisation'' where the population is more or less homogeneously spread over all cities. Of particular interest are cases where there exists a phase transition between the two regimes, as, for example, the migration intensity is increased. In the language of directed polymers, this corresponds to a transition between a pinned glassy phase and an unpinned, {high temperature} 
phase \cite{KPZ,derridaspohn,gueudre}. 

As a generic model for this type of problems, one may consider the following equation, where $x_i(t)$ describes the population of city $i$, or the wealth of individual $i$, or the number of viruses of type $i$, etc. \cite{BouchaudMezardCondensation2000}
\begin{equation} \label{eq:general}
\frac{{\rm d}x_i}{{\rm d}t} = (m_i + \sigma \xi_i(t)) x_i + \sum_{j \neq i} (\varphi_{ij} x_j -  \varphi_{ji} x_i), 
\end{equation} 
for $i=1, \dots, N$. The first term corresponds to multiplicative growth, with mean rate $m_i$ and zero mean, time dependent fluctuations $\xi_i(t)$ (sometimes called ``seascape'' noise \cite{Lassig,Fisher,Kardar2020}). The second term corresponds to migrations/mutations/exchanges, with rates $\varphi_{ij}$ between $j$ and $i$. In some cases, one should add a ``demographic'' noise term $\sqrt{x_i} \eta_i(t)$ and a saturation term $-x_i^2$ \cite{Kardar2020}, but we will not consider them further here. 
The $x_i(t)$ can also be seen as the partition sums of directed polymers of length $t$ with endpoint at site $i$. 
In that context, $m_i$ corresponds to so-called ``columnar disorder'' whereas $\xi_i(t)$ corresponds to point-like disorder \cite{KZDirectP,doussal2025}. 
The long time behaviour of the solution of Eq. \eqref{eq:general} strongly depends on the structure of the graph defined by the migration rates $\varphi_{ij}$, and only partial results are known in the general case. In the one-dimensional, homogeneous case $\varphi_{ij} = \varphi (\delta_{j,i+1} + \delta_{j,i-1})$, $m_i=m$, the free energy corresponds to the height of an interface, which at large scale is
in the universality class of the 
1D KPZ equation. It is known that the population is {\it always localised} at long time
{along an optimal space-time path,} even when $\varphi \to \infty$ \cite{HHZ,KZDirectP}. 
{In presence of
columnar disorder, a third phase with static localisation is generically possible
\cite{NelsonVinokur1993,HwaNelsonVinokur1993}, although less is established
for a general graph. Only recently was its existence demonstrated, in the restricted case of a single columnar defect in one dimension
\cite{Tang1993,HwaNattermann,Basu2014,Soh2017,Krajenbrink2021}.}

Here we want to study the model in {a mean-field, fully connected, limit} i.e. when $\varphi_{ij} = \varphi/N$, $\forall i \neq j$, corresponding to 
\begin{equation} \label{eq:ODE_growth}
\frac{{\rm d}x_i}{{\rm d}t} = (m_i + \sigma \xi_i(t) - \varphi) x_i + \varphi \overline{x}(t); \quad \overline{x}(t) := \frac{1}{N} \sum_i x_i(t)
\end{equation} 
When the mean grow rates are all identical, i.e. $m_i = m$ $\forall i$, and $\xi_i(t)$ are white Gaussian variables (with $\langle \xi_i(t) \xi_j(t')) \rangle = \delta_{ij} \delta(t-t')$), the solution has been worked out in
\cite{BouchaudMezardCondensation2000,Kardar2020}.
When the limit $N \to \infty$ is taken before $t \to \infty$, the population is found to be always delocalised (see also \cite{Medo2009,LiuSerota2018}). The rescaled variables $w_i=x_i/\overline{x}$ reach a stationary distribution at large time, characterized by a power-law tail $w^{-1-\mu}$ for large $w$, with $\mu= 1+ 2 \varphi/\sigma^2$. The case where $m_i$ are heterogeneous has, been much less
investigated
although it corresponds to the realistic case where some cities, individuals, species, etc. have persistent advantages over others (but see \cite{Hur, Zapperi}). This, intuitively, should favour localisation, as we will show below.

\begin{figure}
   \centering
    \includegraphics[width=0.48\textwidth]{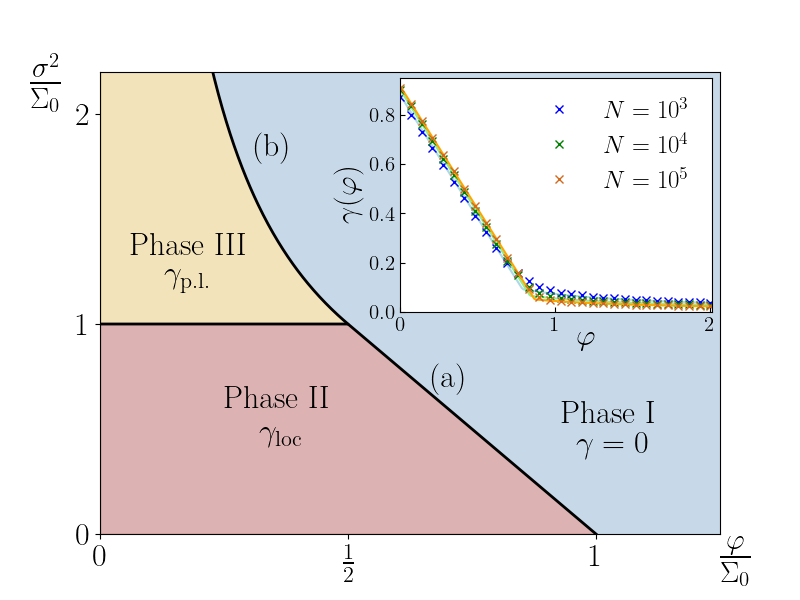}
    \caption{\underline{Main:} Phase diagram predicted for a Gaussian distribution of $m_i$ in presence of noise $\sigma>0$, revealing three distinct phases. Phase I (delocalised with $\gamma=0$) \& II (strongly localised with $\gamma_{\text{loc}} > 0$) are the continuation of the corresponding phases for $\sigma=0$. A new partially localised Phase III emerges, with a growth rate $\gamma_{\rm p.l.} = {\Sigma_0^2}/{2 \sigma^2}-\varphi > 0$, and a power-law distribution of the weights $p_{i}=x_i(t) /\bar x(t)$. 
    The boundaries of phase I are $\varphi_c^{I-II}:= \Sigma_0 - {\sigma^2}/{2}$ and $\varphi_c^{I-III} := {\Sigma_0^2}/{2 \sigma^2}$.
    \underline{Inset:} Plot of the growth rate $\gamma(\varphi)$ evaluated numerically (dots) in the absence of noise $\sigma=0$,
    for various $N$,
    with $m_i$ {i.i.d Gaussian defined in text (with $\Sigma_0=1$).}
    Our prediction (continuous line) is $\gamma =  m_1 - \varphi$ in the strongly localised phase, while in the delocalised phase, one has \cite{footnote1} $\gamma \approx \Sigma_0^2/(2 \varphi \, {\log N}) \to 0$.
    For $N \gg 1$, the transition occurs at $\varphi_c = m_1 - \varphi$ \cite{footnotem1random}. 
    }
    \label{fig:gammamPhases}
\end{figure}

In the first part of this letter, we study the case where $\sigma=0$, i.e. static heterogeneities in growth rates, and show that depending on their distribution $\rho(m)$, one can observe a localisation transition as $\varphi$ decreases below a value $\varphi_c$ that we compute. We then turn to the case $\sigma > 0$ and find a third, partially localised phase, see Fig. \ref{fig:gammamPhases}. We emphasize the strong connections between our model and Derrida's celebrated Random Energy Model \cite{REM}.  

For $\sigma=0$, Eq. \eqref{eq:ODE_growth} can be rewritten in a matrix form $\dot {\bf x} = \mathbb{M} {\bf x}$, with $\mathbb{M}_{ij}= (m_i - \varphi) \delta_{ij} + {\varphi}/{N}$. Noting that $\mathbb{M}$ is a diagonal matrix plus a rank one perturbation, the characteristic polynomial can easily be computed (see SM \cite{SM}) and leads, for $\varphi > 0$, to the following equation for the eigenvalues $\gamma_\alpha$ 
\begin{equation}\label{eq:eq_gamma} 
\frac{1}{N} \sum_i \frac{\varphi}{\gamma_\alpha + \varphi - m_i} = 1.
\end{equation}
Plotting the l.h.s of Eq. \eqref{eq:eq_gamma} versus {$\gamma_\alpha$} shows that these eigenvalues are intertwined with the $m_i$'s, which we will order as $m_1 > m_2 > \ldots > m_N$. The largest eigenvalue $\gamma_1 := \gamma$ corresponds to the asymptotic growth rate of Eq. \eqref{eq:ODE_growth}, {i.e. $\gamma= \lim_{t \rightarrow + \infty} \log \bar x(t)$},
and is such that $\gamma >m_1-\varphi$ and $m_j-\varphi<\gamma_j<m_{j-1}-\varphi$, $j \in [2,N]$.

We now study \eqref{eq:eq_gamma}. For fast redistribution, $\varphi \to +\infty$, the overall growth rate 
converges to the average of the local growth rate over all $i$, i.e one finds (see SM \cite{SM})  
\begin{equation} \label{expansionfixedN} 
\gamma \approx \overline{m} +  \frac{1}{N \varphi} \sum_i (m_i - \overline{m})^2 + O(\varphi^{-2}) \quad , \quad  \overline{m}:=  \frac{1}{N} \sum_i m_i.
\end{equation} 
The first term is expected since the population keeps hopping between all cities, and idiosyncratic effects are averaged out. The correction is positive: decreasing redistribution always {\it increases} the overall growth rate. Indeed { we show}
from \eqref{eq:eq_gamma}, {see SM \cite{SM}}, that $-1 \leq \partial_\varphi \gamma <0$. 

In the opposite limit, for $\varphi \to 0$ and $N$ finite and fixed, 
the fastest growing ``site'' $i=1$ dominates the long term growth.
Indeed in \eqref{eq:eq_gamma} the term $i=1$ dominates, 
leading to $\gamma=m_1-\varphi (1- \frac{1}{N}) + O(\varphi^2)$. One can neglect the contributions of $i \geq 2$ provided $(\varphi/{N}) \sum_{i=2}^N \frac1{m_1-m_i} \ll 1$. In the large $N$ limit, we find that these two regimes are not smoothly connected, i.e. a phase transition occurs.  This is what we discuss next, by focusing on the case where the $m_i$ are independently drawn from a common distribution $\rho(m)$. 

We first consider the case where $\rho(m)$ has a finite upper edge $m_>$, with $\rho(m > m_>)=0$,
such that the Stieltjes transform $G(z)= \int {\rm d}m \rho(m) (z-m)^{-1}$ is well defined for $z$ outside the support of $\rho(m)$. Defining the so-called $R$-transform of $\rho$ as $R(z)=G^{-1}(z) - 1/z$, Eq. \eqref{eq:eq_gamma} can be simply rewritten for large $N$ as 
\begin{equation}\label{eq:R-transform}
   {G(\gamma+\varphi) = \frac{1}{\varphi} \quad \Rightarrow} \quad \gamma = R\left(\frac{1}{\varphi}\right). 
\end{equation}
The $R$-transform is of primary importance in the context of free random matrices, and the coefficient of its series expansion in $z$ are called free cumulants.
The exact shape of $R(z)$ is known in several cases. For example, when $\rho(m)=\sqrt{4 \Sigma^2 - m^2}/2\pi \Sigma^2$  (Wigner's semi-circle), one has $R(z)=\Sigma^2 z$ and $m_>=2 \Sigma$. For the arc-sine distribution $\rho(m)=1/(\pi \sqrt{1-m^2})$, $R(z)=(\sqrt{1+z^2}-1)/z$ and $m_>=1$. In the first case, $\rho(m)$ behaves as $({m_>-m})^{1/2}$ close to the upper edge, whereas in the second case $\rho(m)$ behaves as $({m_>-m})^{-1/2}$, leading to very different conclusions that we will generalize below \cite{footnotewishart}.  Indeed, Eq. \eqref{eq:R-transform} tells us that for the Wigner semi-circle the growth rate $\gamma$ is equal to $\Sigma^2/\varphi$, which exceeds $m_>=2 \Sigma$ when $\varphi < \Sigma/2$. But this cannot be since $m_>$ is the maximum possible growth rate. This is because Eq. \eqref{eq:R-transform} only holds when $\gamma + \varphi > m_>$ (see Eq. \eqref{eq:eq_gamma}), or $\varphi > \Sigma$. When $\varphi < \varphi_c = \Sigma$, $\gamma + \varphi$ actually sticks to the maximum value $m_>=2 \Sigma$. As we shall see below, this situation corresponds to a {\it localisation transition}, for which a finite fraction of the population is actually found on site $i=1$ and benefits from a maximum growth rate, despite the ceaseless ``hopping around'' of each individual at rate $\varphi$, resulting in a global growth rate given, for large $N$, by $\gamma = m_> - \varphi$. 

The arc-sine distribution leads to $\gamma=\sqrt{1+\varphi^2} - \varphi$ which now is always smaller than $m_>=1$ when $\varphi > 0$ (and indeed $\gamma + \varphi > 1$). In this case, the population is always delocalised over all sites. 

The above results generalize to { a larger class} of distributions with finite support, and with $\rho(m) \sim (m_> - m)^{\psi -1}$ close to the upper edge, with $\psi=3/2$ (resp. $1/2$) for the Wigner (resp. arc-sine) case. Whenever $\psi > 1$, there is a localisation transition at $\varphi_c = 1/G(m_>)$. For $\varphi > \varphi_c$, the growth rate $\gamma$ is the unique solution of Eq. \eqref{eq:R-transform}, whereas for $\varphi < \varphi_c$, the population localises and the growth rate is given again by $\gamma = m_> - \varphi$. 
In the case $\psi \leq 1$ on the other hand, $G(m_>)=\infty$ and the population remains delocalised as soon as $\varphi > 0$. 
The resulting $\gamma(\varphi)$ is checked numerically in Fig.S3 of the SM \cite{SM}.

Note that, interestingly, an equation similar to  Eq. \eqref{eq:R-transform} appeared very recently \cite{Gueneau2025} in a model of a 1D Brownian particle that resets
its diffusion coefficient $D$ with rate $r=\varphi$, to a random value distributed according to $\rho(D)$.
The analog of our transition for $N \to +\infty$ first, i.e. for an infinite pool of values $D_i$, was observed in the large time limit by the authors of \cite{Gueneau2025}. For a finite pool $\{ D_i\}_{i \in [1,N]}$, we thus predict the same condensation transition (dominance of $D_{\rm max}$) as obtained here, 
see Section C in SM \cite{SM}.

Why are we talking about localisation when $\varphi < \varphi_c$? Mathematically, what happens is akin to the theory of Bose condensation, when a finite fraction of molecules ``condense'' into the lowest energy state \cite{Huang}. In order to see this, we first note that at large times, such that $(\gamma - \gamma_2) t \gg 1$ and fixed $N$, one has, from Eq. \eqref{eq:ODE_growth},
{ for any given set of $m_i$}

\begin{equation}\label{eq:ODE_sol}
x_i(t) \underset{t \to \infty}{\sim}  \frac{\varphi}{\gamma + \varphi - m_i} \overline{x}(t), \; \text{with} \; \;  \overline{x}(t) \propto e^{\gamma t}.
\end{equation} 
{Defining $p_i(t)=x_i(t)/\sum_j x_j(t)$ the probability of occupying site $i$, we see that
asymptotically it behaves as $p_i \simeq {\varphi}/(N({\gamma + \varphi - m_i}))$,}
and Eq. \eqref{eq:eq_gamma} simply reads $\sum_i p_i=1$. Now, suppose that $0 < \varphi < \varphi_c$ 
and expand $\gamma$ as 
{ $m_1 - \varphi(1 - \epsilon)$}
with ${ \epsilon  < 1/N} \ll 1$ to be determined. Inserting in Eq. \eqref{eq:eq_gamma} we obtain $p_1 = 1/(N \epsilon) = 1 - X_N$ with (see \cite{SM})
\be \label{sum} 
{ X_N :=} 
\frac1N \sum_{j=2}^N \frac{\varphi}{\varphi \epsilon + m_1-m_j} {\overset{N \gg 1}{\simeq} } \varphi \int_0^{+\infty} {\rm d}y  \frac{\rho(m_1-y)}{\epsilon \varphi + y} 
\ee 
Provided the last integral converges when $\epsilon \to 0$, we recognize precisely $G(m_1) \to_{N \to \infty} G(m_>) = 1/\varphi_c$. Hence:
\be \label{p1phi}
p_1 = \frac{1}{N \epsilon} = \left(1 - \frac{\varphi}{\varphi_c}\right),
\ee
meaning that indeed, when ${\varphi}<{\varphi_c}$ a finite fraction of the population is in a single ``state'' 
\cite{occupation}
, as in the Bose condensation problem.
This fraction vanishes continuously as ${\varphi} \to {\varphi_c}^-$, and remains zero in the delocalised phase, see Fig. \ref{fig:gammam} in the SM \cite{SM}.


We now turn to the case where $\rho(m)$ has an infinite support, with $m_> = + \infty$. In order to be specific, we consider the following family of generalized Gaussians: $\rho(m) = \exp{(-(m^2/2 \Sigma^2)^b)}/Z_b$ where $b > 0$ and $Z_b$ an appropriate normalisation. The Gaussian case corresponds to $b=1$, whereas $b=1/2$ is the double Laplace distribution. From extreme value theorems, we know that the largest value of $m$ after $N \gg 1$ independent draws is 
{ $m_1=m_1(N) \sim \Sigma \sqrt{2 (\log N)^{1/b}} $}. To keep the maximum growth rate finite as $N$ goes to infinity, we choose {$\Sigma = \Sigma_0/\sqrt{2 (\log N)^{1/b}}  $,} such that $m_1 \approx \Sigma_0$. 

We now again assume that $\gamma = m_1 - \varphi(1-\epsilon)$. Looking at the integral in Eq. \eqref{sum}, one can convince oneself that it is dominated by the saddle point at $y = m_1$ (i.e. by the typical rates $m_i \approx 0$, see \cite{SM} for details), such that $X_N \to \varphi/\Sigma_0$. Hence, for $\varphi < \varphi_c = \Sigma_0$ the system is localised with $p_1 = 1 - \varphi/\varphi_c$, and becomes delocalised above $\varphi_c$. But now, because of our scaling of $\Sigma$ with $(\log N)^{1/2b}$, we find that in the limit $N \to \infty$
\be \label{eq:gammaF}
\gamma = \begin{cases} 
\Sigma_0 - \varphi \quad , \quad \varphi < \varphi_c=\Sigma_0 \\
0 \quad , \quad \varphi > \varphi_c 
\end{cases}
\ee 
This prediction is tested in Fig. \ref{fig:gammamPhases}. 
If one had chosen to keep $\Sigma$ constant, independent of $N$, then the localisation transition would take place for a logarithmically increasing value of $\varphi_c \sim
\Sigma (\log N)^{1/2b}$. 

We now turn to the very interesting case where static, heterogeneous growth rates $m_i$ compete with time-dependent fluctuations $\xi_i(t)$, see Eq. \eqref{eq:ODE_growth}. 
When $m_i=m$, we know from previous work that the population is always delocalised when $N \to \infty$ before $t \to \infty$
\cite{footnoteHerfindal}.
The case $N=2$ can be solved exactly, as shown in \cite{bouchaudtax}. What happens when we keep $N$ large but finite and take $t$ to infinity, with both heterogeneous static growth rates $m_i$ and time-dependent fluctuations? \cite{footnoteGabaixIneq} 

In order to investigate this question, let us revisit the static problem with Gaussian $m_i$'s (i.e. $b=1$ henceforth) from another point of view, establishing a strong connection with Derrida's Random Energy Model \cite{REM}. Consider the following auxiliary problem of estimating
$Z_\Omega = \sum_{i=1}^N e^{S \Omega_i}$, 
where $\Omega_i$ are independent $N(0,1)$ random variables, and both $S$ and the number terms in the sum $N$ tend to infinity. Then, Derrida's result is that there is a phase transition for $S = S_c = \sqrt{2 \log N}$ between a self-averaging phase where $Z_\Omega$ is given by the law of large numbers, and a ``frozen'' phase where $Z_\Omega$ is dominated by a finite number of particularly large terms. More precisely (see \cite{benarous}),
\be \label{eq:REM}
Z_\Omega \approx 
\begin{cases}
N e^{S^2/2} &\qquad  S < S_c  \\
e^{S \sqrt{2 \log N}} \chi^{(\mu)} &\qquad  
S > S_c,
\end{cases}
\ee 
where $\chi^{(\mu)}$ is a (totally asymmetric) L\'evy stable random variable of index $\mu=S_c/S < 1$. Now, working with the It\^{o} convention, the explicit solution of Eq. \eqref{eq:ODE_growth} is
\bea \label{eq:ODE_growth2}
x_i(t) = x_0  \psi(t) 
+ \varphi \psi(t) \int_0^t {\rm d}t' \,   \overline{x}(t')/\psi(t'),
\eea
where $\psi(t):=\exp[{(m_i - \varphi - \frac{\sigma^2}{2}) t + \Xi_i(t)}]$ and $\Xi_i(t):=\sigma \int_0^t {\rm d}u \, \xi_i(u)$ is a Brownian motion of variance $\sigma^2 t$. Summing this equation over $N$, the second term on the right hand side of this equation reads 
\be \label{eq:Z}
Z(t) \approx \varphi \Gamma e^{\gamma t} \int_0^t {\rm d}u \, e^{-(\varphi + \gamma +\frac{\sigma^2}{2})u} \sum_{i=1}^N e^{S(u) \Omega_i},
\ee 
where we have assumed, as above, that for large $u$, $\overline{x}(u) \approx \Gamma e^{\gamma u}$, and where $S$ is the standard deviation of the Gaussian variable $m_i u + \Xi_i(u):=S \Omega_i$, given by $S^2(u)=\Sigma^2 u^2 + \sigma^2 u$. Now, the REM result Eq. \eqref{eq:REM} tells us that we should cut the integral over $u$ in Eq. \eqref{eq:Z} at $u_c$ such that $S(u_c)=\sqrt{2 \log N}$, i.e.
\bea \label{eq:Z2}
Z(t) &\approx& \varphi N \Gamma e^{\gamma t} \left[\int_0^{u_c} {\rm d}u \, e^{-(\varphi + \gamma +\frac{\sigma^2}{2})u + \frac{S^2(u)}{2}} \right. \\ \nonumber &+& \left. \frac1N \int_{u_c}^t {\rm d}u \, e^{-(\varphi + \gamma +\frac{\sigma^2}{2})u + S(u) \sqrt{2 \log N}} \chi_{u}^{(\mu)} \right],
\eea
with the self-consistent condition that $Z(t) \approx N \Gamma e^{\gamma t}$ at large times, coming from the fact that $\sum_i x_i(t)= N \overline{x}(t)$. \cite{footnotefluct}

\begin{figure}
   \centering
    \includegraphics[width=0.55\textwidth]{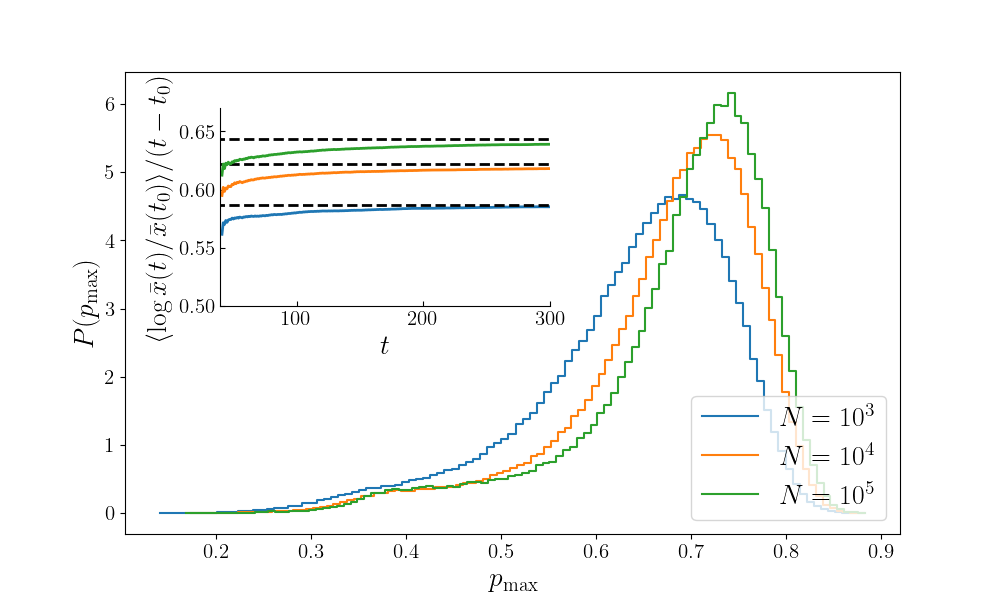}
    \caption{Main: Plot of the probability distribution of $p_{\rm max}=\max_i x_i/\sum_j x_j$ for various $N$ for $\varphi=0.2$ and $\sigma=0.3$. It shows that the system is localised for these values of the parameters. Inset: corresponding effective growth rates measured as $ \langle \log{\frac{\bar x(t)}{\bar x(t_0)}} \rangle/(t-t_0)$ 
    for $t_0=40$ (continuous line), compared to the prediction $\gamma= m_1 - \varphi - \frac{\sigma^2}{2}$ {(dashed lines)}.
    The value of $m_1=m_1(N)$ converges to $\Sigma_0=1$ with logarithmic corrections.\cite{footnotefigpmax}}
    \label{fig:pQ1sig}
\end{figure}

The whole discussion now is to determine which integral is dominant in the long time limit $t \to \infty$. Let us scale the variance $\Sigma^2$ of the growth rate $m_i$ as previously, namely $\Sigma^2 = \Sigma_0^2/2 \ell$ with $\ell:=\log N$ henceforth. We then find
\be
u_c = \ell \tau_c, \quad \tau_c = \frac{1}{\Sigma_0^2} \left(
\sqrt{\sigma^4 + 4 \Sigma_0^2} - \sigma^2 \right)
\ee
The second integral in Eq. \eqref{eq:Z2} then reads, after a change of variable $u \to v u_c$:
\be \label{eq:I2}
I_2 = \frac{\varphi \ell \tau_c}{N} \int_1^\infty {\rm d}v \, e^{- \ell f_\gamma(v)}.   
\ee
with $f_\gamma(v)=(\varphi + \gamma + \frac{\sigma^2}{2}) v \tau_c - \sqrt{2 \sigma^2 \tau_c v + \Sigma_0^2 \tau_c^2 v^2}$ and where we have neglected the random $\chi$ term, of order unity. Due to the $1/N$ factor in front of $I_2$, one may naively assume that it is always negligible compared to the first integral in Eq. \eqref{eq:Z2}, $I_1$. 
However, $f_\gamma(v)$ can be negative and $I_2$ is of order one when $f_\gamma(v)$ admits a minimum $v^\star$ s.t. $f_\gamma(v^\star) = -1$. This is achieved when $\gamma=\gamma_{\textrm{p.l.}}$ with 
\be
\gamma_{\textrm{p.l.}}=\frac{\Sigma_0^2}{2 \sigma^2}-\varphi \quad , \quad v^\star= \frac{2 \sigma^2}{(\sigma^4-\Sigma_0^2) \tau_c}
\ee
which is possible only when $\sigma>\Sigma_0^2$ and $\varphi < \varphi_c^{I-III} := {\Sigma_0^2}/{2 \sigma^2}$. This corresponds to the low temperature, partially localised phase of the REM. The equivalent of the rescaled temperature for the REM is $\mu={S_c}/{S}(u^\star) < 1$ where $u^\star$ is the saddle-point of $I_2$. Hence
\be \label{mufrozen}
\mu= 1- \frac{\Sigma_0^2}{\sigma^4} 
\ee
Using the mapping to the REM, we predict that the distribution of $p_{\max}=\max_i x_i(t) /\bar x(t)$  decays as $\sim 1/p_{\max}^{1+\mu}$, as checked in Fig.S3 of the SM \cite{SM}. 
For $\sigma < \Sigma_0$, 
$I_2$ starts diverging when $\varphi + \gamma + \sigma^2/2 \to \Sigma_0^+$, giving a contribution $\sim [N (\varphi + \sigma^2/2 + \gamma - \Sigma_0)]^{-1}$ which becomes of order one. This is the analogue of the calculation of $p_1$ above, which becomes of order unity in the localised phase. Hence in that phase, $\gamma = \gamma_{\rm loc} = \varphi_c^{I-II} - \varphi$ with $\varphi_c^{I-II}:= \Sigma_0 - {\sigma^2}/{2}$. { We test this prediction in the inset of Fig. \ref{fig:pQ1sig} while the main plot shows the distribution of $p_{\rm max}=\max_i p_i$. As $N$ becomes large, $p_{\rm max}$ is $O(1)$ with
fluctuations of the same order. Note that $\gamma_{\rm loc}=\Sigma_0-\varphi-\frac{\sigma^2}{2} = \lim_{N\to +\infty} m_1(N)-\varphi-\frac{\sigma^2}{2}$ corresponds to the asymptotic growth rate $m_1(N)$ of the site where a finite fraction of the population concentrates. This allows us to predict finite size corrections. 
}

We have so far neglected the contributions of $I_1$ which are indeed negligible as long as $\gamma_{\rm p.l.}$ or $\gamma_{\rm loc} >0$, otherwise $I_2 \ll I_1$. After the change of variable, $I_1$ then reads
\be \label{eq:I1}
I_1 = \varphi \ell \tau_c \int_0^1 {\rm d}v \,  e^{- \ell \left((\varphi + \gamma) v \tau_c - \Sigma_0^2 \tau_c^2 v^2/4 \right)}, 
\ee 
which, provided $\varphi + \gamma > \Sigma_0^2 \tau_c/4$, 
is dominated for large $\ell$ by the vicinity of $v=0$, so that $I_1 \approx \varphi/(\varphi + \gamma)$ \cite{footnotecontI1}.
In that case the self-consistent condition $Z(t) = N \Gamma e^{\gamma t}$ is equivalent to $I_1=1$, which 
leads to $\gamma=0$ for the global growth rate, 
as we found above in the delocalised phase (see Eq. \eqref{eq:gammaF}).

Hence, our REM analysis predicts a rather rich phase diagram with three phases, summarized in Fig. \ref{fig:gammamPhases} in the plane $(\varphi/\Sigma_0, \sigma^2/\Sigma_0)$. When $\sigma^2=0$, we recover the results above, Eq. \eqref{eq:gammaF}, using a different framework, which emphasizes the deep connections of our problem with the REM. But as $\sigma^2$ becomes large enough, the localised phase observed for $\varphi < \Sigma_0$ disappears and becomes delocalised with $\gamma = 0$. When $0 < \varphi < \Sigma_0/2$, the system first goes through a partially localised phase with $\gamma=\gamma_1$ before becoming fully delocalised. In particular, our result recovers the fact that there is no localisation for large $N$ when $\Sigma_0=0$ \cite{BouchaudMezardCondensation2000}. [A more refined analysis shows however that there is a crossover between a localised region and delocalised region when $\log N \sim \sigma^2/2 \varphi$ \cite{footnotecrossover}, which thus becomes relevant when $\varphi \to 0$.] 
We have performed numerical checks of these results. We have clear evidence
of a localised phase with the predicted value for $\gamma_{\rm loc}$, see Fig. \ref{fig:pQ1sig}
in the SM \cite{SM}. We also confirm the presence of the partially localised phase, although the precise determination of $\gamma_1$ is made difficult by $1/\log{N}$ corrections. Note that our model corresponds to the mean-field limit of the model considered in \cite{Hur}, which did not mention the existence nor the nature of the three phases reported here.
Many directions are left to explore, such as the influence of heterogeneities in the variance $\sigma^2$ across individuals or of the network topology on the phase diagram of Fig. \ref{fig:gammamPhases}.


In conclusion, heterogeneous mean growth rates lead to much stronger level of inequalities than volatile but homogeneous growth rates. In the latter case, uniform redistribution, however small, always leads to moderate inequalities, characterised by fat tails but with a finite mean \cite{BouchaudMezardCondensation2000}, as we mentioned in the introduction. In contrast, when mean growth rates are heterogeneous, a minimum level of redistribution $\varphi_{\text{c}}$ is needed to ensure that extreme concentration does not occur. However, this comes at a cost since the corresponding global growth rate $\gamma$ is substantially reduced from its maximum value, dominated by outliers. When both effects are present, a sufficient level of temporal fluctuations in growth rates can mitigate extreme concentrations (see Fig. \ref{fig:gammamPhases}). In the context of wealth inequalities, this means that persistent differences of ``skill'' or ''rents'' (i.e. different $m_i$s) can only lead to extreme concentrations (or ``oligarchies'' as coined in \cite{Boghosian2019}), except if a sufficient wealth tax rate $\varphi > \varphi_c$ is introduced as a redistribution tool. This is still the case when success is the result of a mix of ``skill'' and ``luck'' \cite{Pluchino, Hur}, albeit with a lower value of $\varphi_c$. 
When the luck component $\sigma^2$ is large enough (i.e. $> \Sigma_0$), a smaller wealth tax rate $\varphi \geq \Sigma_0^2/2 \sigma^2$ is enough to prevent extreme concentration. In our highly stylized setting, accidental ``bad streaks'' are enough to avoid a fraction of total wealth being concentrated in the hands of a few individuals.  Even in the partially localised phase, interestingly, the ``happy few'' are not fixed in time: individuals may get lucky one day or another. \cite{footnoteconcl}

{\bf Acknowledgments}. PLD acknowledges support from ANR Grant ANR-23-CE30-0020-01 (EDIPS).
This work was performed using computational resources from LPTMS.

\newpage






\clearpage

\newpage

.

\newpage

\begin{widetext}


\setcounter{equation}{0}
\setcounter{figure}{0}
\renewcommand{\thetable}{S\arabic{table}}
\renewcommand{\theequation}{S\thesection.\arabic{equation}}
\renewcommand{\thefigure}{S\arabic{figure}}
\setcounter{secnumdepth}{2}


\begin{center}
	\textbf{\large  A mean-field theory for heterogeneous random growth with redistribution
		\\
		[.3cm] -- Supplemental Material --} \\
	[.4cm] Maximilien Bernard,$^{1,2}$ Jean-Phillipe Bouchaud,$^3$ and  Pierre Le Doussal $^1$\\[.1cm]
	{\itshape $^1$Laboratoire de Physique de l'Ecole Normale Sup\'erieure, \\ 
       CNRS, ENS and PSL Universit\'e, Sorbonne Universit\'e, \\ Universit\'e Paris Cit\'e, 24 rue Lhomond, 75005 Paris, France} \\
{\itshape $^2$LPTMS, CNRS, Univ. Paris-Sud, Universit\'e Paris-Saclay, 91405 Orsay, France} \\
	{\itshape $^3$ Capital Fund Management, 23 rue de l'Université, 75007 Paris, France \& Académie des Sciences, 75006 Paris, France}\\
\end{center}







\section{Properties of $\gamma$ versus $\varphi$, Free Cumulants} 

Using the matrix determinant lemma, the characteristic polynomial is simply
\be \label{det} 
\det (z I - \mathbb{M} ) = \prod_{i=1}^N (z- m_i + \varphi) \times \left(1 - \frac{\varphi}{N} \sum_{i=1}^N \frac{1}{z- m_i + \varphi} \right)
\ee 
Hence for $\varphi > 0$ and if all $m_i$ are distinct (assumed here), the $N$ eigenvalues $\gamma_\alpha$ of $M$ are the $N$ roots of \eqref{eq:eq_gamma}, with $\gamma=\gamma_1=\max_\alpha \gamma_\alpha$.

Next we show that $\gamma$ decreases with $\varphi$.
Let us define $\rho_N(m)$ the empirical distribution 
\be 
\rho_N(m)=\frac{1}{N} \sum_i \delta(m-m_i) \label{empirical} 
\ee
for a given set of $m_i$.
Taking derivatives of \eqref{eq:eq_gamma} one obtains
\be 
-1 < \partial_\varphi \gamma = - \frac{ \int dm  \frac{\rho_N(m)}{(\gamma + \varphi - m)^2 } - (\int dm  \frac{\rho_N(m)}{(\gamma + \varphi - m) } )^2 }{
\int dm  \frac{\rho_N(m)}{(\gamma + \varphi - m)^2 } } <0 
\ee 

Let us now consider the expansion in $1/\varphi$. 
The coefficients of the series expansion in $z$ of the $R$ transform, $R(z)= \sum_{q \geq 0}  \kappa_{q+1} z^q$, 
define the free cumulants $\kappa_q$
associated to the density $\rho(m)$. From \eqref{eq:R-transform} we 
thus obtain the large $\varphi$ expansion
\be \label{expansion} 
\gamma= R(1/\varphi)= \sum_{q \geq 0}  \kappa_{q+1} \varphi^{-q} 
\ee 
The free cumulants can be expressed in terms of the moments
$\mu_q= \int dm m^q \rho(m)$ of the density $\rho(m)$. For $q=1,2,3$
the $\kappa_q$'s are identical to the usual cumulants, with $\kappa_1=\mu_1$,
$\kappa_2=\mu_2-\mu_1^2$. For $q \geq 4$ they are different,
see e.g. Appendix B in \cite{Gueneau2025} for some explicit formula. 
It is important to note that \eqref{expansion} remains
true for finite $N$ and for a given set of $m_i$, where the $\kappa_q$ are then the 
free cumulants associated to the empirical distribution $\rho_N$ in \eqref{empirical}, and are
expressed by the same formula in terms of the moments $\mu_q = \frac{1}{N} \sum_i m_i^q$.
This leads to the result \eqref{expansionfixedN} in the text.
An important remark is that the series \eqref{expansion} is
useful/valid only in the delocalised phase, as non-perturbative
corrections become important near the transition to the localised phase.

\section{ Evaluation of $X_N$}  

We need to evaluate $X_N$ defined in \eqref{sum} for fixed $m_1=\max m_j$,
at large $N$ for $\epsilon=\tilde \epsilon/N$ with $0<\tilde \epsilon<1$.  
Consider the case of i.i.d. variables with PDF called here $p_N(m)$ (since it may also
depend on $N$, e.g. in the Gaussian case). Conditioning on $m_1$ it is easy to see that the variables $m_j$
for $j \geq 2$ are still i.i.d variables with PDF
$p_N(m|m_1) = \frac{p_N(m) \theta(m<m_1)}{P_{<}(m_1)}$.
Here $P_{<}(m_1)=1- P_{>}(m_1)$ 
where $P_{>}(m) = \int_m^{+\infty} p_N(m)$, and we know
that the CDF of $m_1$ at large $N$ takes the form
$Q_{<,N}(m_1) \sim e^{- N P_{>}(m_1)}$. Hence typically
$P_{>}(m_1) = O(1/N)$ and we can approximate
$p_N(m|m_1) \simeq p_N(m) \theta(m<m_1)$.

Now at large $N$ we can estimate $X_N$ (at fixed $m_1$) by its mean over the $m_j$, $j \geq 2$, conditional to the value of $m_1$, which we denote $\bar X_N$ (which at finite $N$ is still a random quantity, because $m_1$ fluctuates)
\be \label{formXbN} 
X_N \simeq \bar X_N =  \varphi  \int_0 d\mu  \frac{p_N(m_1-\mu)}{\epsilon \varphi + \mu} 
\ee 

{\it Gaussian case}. Consider the Gaussian example, $p_N(m)= \frac{1}{\sqrt{2 \pi } \Sigma_N} e^{- m^2/(2 \Sigma_N^2)}$.
At large $N$, $m_1$ is distributed so that $N P_{>}(m_1)=e^{-u}$
where $u$ is a Gumbel variable with CDF $e^{-e^{-u}}$. 
Using that $N P_{>}(m_1) \simeq \frac{N \Sigma_N}{m_1 \sqrt{2 \pi}} e^{- m_1^2/(2 \Sigma_N^2)}$,
one recovers the standard result
\be \label{standard} 
m_1 \simeq \Sigma_N \sqrt{2 \log N} (1 + \frac{u + c_N }{2 \log N} ) \quad , \quad c_N = - \frac{1}{2} \log (4 \pi \log N)
\ee 
As in the text we choose $\Sigma_N = \Sigma_0/(\sqrt{2 \log N})$ so that $m_1 \simeq \Sigma_0$ to leading order.
From \eqref{formXbN} we have 
\be \label{cltg2}
\bar X_N = \varphi \int_0^{+\infty}  d\mu   \frac{\frac{1}{\sqrt{2 \pi } \Sigma_N} e^{- (m_1-\mu)^2/(2 \Sigma_N^2)}}{ \epsilon \varphi + \mu }
\ee 

This integral can be evaluated by the saddle point method, the saddle point being
at $\mu=m_1$, corresponding to the typical $m_i \approx 0$. Using the normalization condition of the Gaussian, and $\epsilon = O(1/N) \ll 1$, one readily finds that at large $N$
\be 
\bar X_N \simeq \varphi \int^{m_1}  dm   \frac{\frac{1}{\sqrt{2 \pi } \Sigma_N} e^{- m^2/(2 \Sigma_N^2)}}{ m_1  }  
\simeq \frac{\varphi}{\Sigma_N \sqrt{2 \log N} } = \frac{\varphi}{\Sigma_0}  \label{evaluateX} 
\ee 
does not fluctuate any more, and its limit is as given in the text. 

It is interesting to note that the region $\mu$ near $0$ in \eqref{formXbN} (which corresponds to
the secondary maxima, or smallest gaps $g_j=m_1-m_j$) contains a logarithmic divergence,
which is cutoff by the term $\epsilon \varphi$ (recall that $\epsilon=\tilde \epsilon/N$ with $\tilde \epsilon=O(1)$). The contribution
of this region to $\bar X_N$ (for a fixed value of $m_1$ or equivalently of the Gumbel variable $u$ from \eqref{standard}) can be estimated as follows
\be
\bar X_N|_{\mu \approx 0}  \simeq \varphi \frac{e^{- m_1^2/(2 \Sigma_N^2)}}{\sqrt{2 \pi } \Sigma_N}  \int^{m_1}  dm   \frac{1}{ \epsilon \varphi + m_1 - m }  
 \simeq \varphi \frac{m_1 e^{-u}}{N \Sigma_N^2}  \int_0^{m_1}  d\mu   \frac{1}{ \epsilon \varphi + \mu }  
\simeq \varphi \frac{1}{N} e^{-u} \frac{2 \log N}{\Sigma_0} \log (N/\varphi \tilde \epsilon) 
\ee 
which is negligibly small compared to \eqref{evaluateX}. If one includes larger values of $\mu$ in the integral in \eqref{formXbN} the numerator increases rapidly and one 
eventually gets dominated by the saddle point. 

Note that one can also estimate $X_N$ heuristically using the properties of the gaps. 
Indeed for fixed $n$ and large $N$ they behave as (neglecting their fluctuations)
\begin{equation}
g_n= m_1-m_n \approx \Sigma_N \left[ \sqrt{2\log N} - \sqrt{2\log N/n}\right].
\end{equation} 
Since one has that $\varphi \epsilon = O(1/N) \ll g_2$, the above sum can then be estimated, for large $N$ as
\begin{equation}\label{eq:eq_gamma3} 
\bar X_N \simeq \frac{\varphi}{\sqrt{2}\Sigma_N N} \int_2^N {\rm d}n \frac{1}{\sqrt{\log N} - \sqrt{\log (N/n)}} \approx 
\frac{\varphi}{\Sigma_N \sqrt{2\log N}} = \frac{\varphi}{\Sigma_0} 
\end{equation}
in agreement with the previous result \eqref{evaluateX} (it is again dominated by the typical $n \sim N$)

{\it Distributions with an upper edge}. Consider now the case $p_N(m)=p(m)=\psi (1-m)^{\psi-1} \theta(0<m<1)$, i.e.
$P_{>}(m) \simeq (1-m)^\psi$, with for simplicity $m_>=1$. At large $N$, the random variable $m_1$ can be written 
as $m_1 = 1- y/N^{1/\psi}$ where $y$ is a positive random variable with 
${\rm Prob}(y>Y) = e^{- Y^\psi}$. Consider the case $\psi>1$ where there is a localization transition.
The mean $\bar X_N$, conditional to $m_1$ (i.e. conditional to $y$) reads
\bea \label{integral}
&& \bar X_N \simeq  \varphi  \int_0^{m_1} d\mu  \frac{p(m_1-\mu)}{\epsilon \varphi + \mu} 
= \varphi  \psi \int_0^{1-y/N^{1/\psi}} d\mu  \frac{ (y/(N)^{1/\psi} +\mu)^{\psi-1}}{\tilde \epsilon \varphi/N + \mu}
\simeq  \varphi \psi  \int_0^{1} d\mu   \mu^{\psi-2} = \frac{\varphi \psi}{\psi-1} = X_\infty
\eea 
which also reads $\bar X_\infty= \varphi \int dm \frac{p(m)}{m_e-m}$.
This estimate is not trivial and states that one can take the two cutoffs to zero.

\begin{figure}
   \centering
    \includegraphics[width=0.7\textwidth]{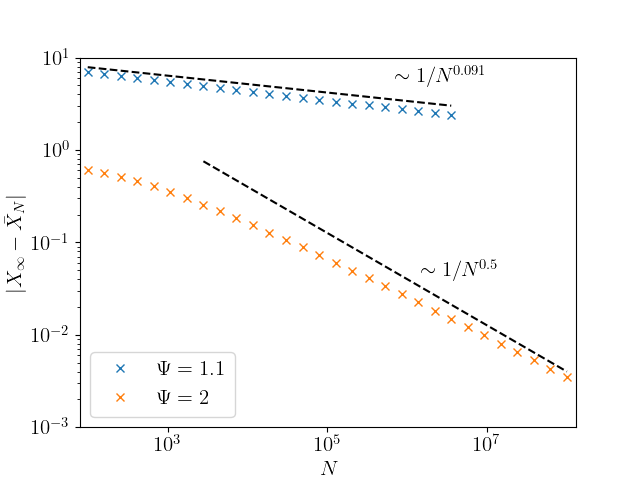}
    \caption{Numerical evaluation of $\bar{X}_N$ compared to the predicted value $X_\infty$ for $\Psi = 1.1$ and $\Psi = 2$ for several values of $N$. The parameters are set as $\varphi = 1$, $y = 1$, and $\tilde{\epsilon} = 1$. The results demonstrate that the difference between $\bar{X}_N$ and $X_\infty$ vanishes in the large $N$ limit,with a decay given by the lower cutoff $\sim N^{1-1/\Psi}$ (indicated by the black dashed line).}
    \label{fig:evint}
\end{figure}

Indeed the integral on scales $\mu = \tilde \mu/N < c/N^{1/\psi}$ with $c \ll 1$ is 
of order $\varphi \psi  \int_0^{c N^{1- \frac{1}{\psi}}} \frac{d \tilde \mu}{\tilde \epsilon \varphi + \tilde \mu} y^{\psi-1} N^{- (1 - \frac{1}{\psi})} \sim 
N^{- (1 - \frac{1}{\psi})} \log N $,
which decays to zero. On scales $1/N \ll \mu = \hat \mu/N^{1/\psi} \ll 1$ 
one forms the difference between $\bar X_N$ and its expected asymptotics $X_\infty$
\be
\bar X_N - X_\infty = \varphi \psi \int_{O(N^{-1})}^1 d\mu \mu^{\psi-2} \left( (1 + \frac{y}{\mu N^{1/\psi}})^{\psi-1}-1 \right) 
\simeq  \varphi \psi N^{\frac{1}{\psi} -1} 
\int_{O(N^{1/\psi-1})}^{N^{1/\psi}} d\hat \mu  \hat \mu^{\psi-2} \left( (1 + \frac{y}{\hat \mu})^{\psi-1}-1 \right)  
\ee 
where in the last integral we have written $\mu = \hat \mu/N^{1/\psi}$. For $1<\psi<2$ this integral is convergent at large $\hat \mu$
and the total result decays as $\sim N^{- (1- \frac{1}{\psi})}$ (up to a logarithmic factor originating from the lower bound, as noted above).
For $\psi>2$ one must use the upper bound $\hat \mu \sim N^{1/\psi}$ leading to a
decay $N^{-1/\psi}$.

To check these estimates we have performed a numerical
estimate of the integral in \eqref{integral}. The results are shown 
in Fig.\ref{fig:evint} 

\section{ Growth and resetting}

As mentioned in the text, for $\sigma=0$, our population growth model 
can also be seen a resetting problem, 
where a particle randomly resets its growth rate, $m_i \to m_j$ through a jump
from site $i$ to $j$, at rate
$r=\varphi$. To pursue the analogy, the quantity which undergoes 
growth in the model of Ref. 
 \cite{Gueneau2025}, is $\mathbb{E} [ e^{q X(t) }]$ where
 $X(t)$ is a 1D Brownian motion which resets
its diffusion coefficient $D$ with rate $r$. The
growth rate there, $\Psi(q)= \lim_{t \to +\infty}  \frac{1}{t} \log \mathbb{E} [ e^{q X(t) }]$,
is thus the analog of $\gamma$ here, with $m_i=D_i q^2$. 


Let us further remark that for $\sigma=0$ it is easy to solve \eqref{eq:ODE_growth2} 
using Laplace transforms (LT). Let us define 
$\hat x_i(s)= \int_0^{+\infty} dt e^{-s t} x_i(t)$
and $\hat x(s)= \int_0^{+\infty} dt e^{-s t} \bar x(t)$.
Taking the LT of \eqref{eq:ODE_growth2} one obtains (for $s+\varphi> m_1$)
\be 
\hat x_i(s)= \frac{x_0 + \varphi \hat x(s)}{s+\varphi- m_i} 
\ee 
Summing, one obtains
\be 
\hat x(s)  = G_N(s+\varphi) (x_0 + \varphi \hat x(s) ) 
\ee 
where $G_N(s)=\frac{1}{N} \sum_i \frac{1}{s - m_i}$ is the Stieltjes transform of $\rho_N(m)$
in \eqref{empirical}, 
leading to
\be 
\hat x(s) = x_0 \frac{G_N(s+\varphi)}{1- \varphi G_N(s+\varphi)}
\ee 
One recovers that there are simple pole singularities at $s=\gamma_\alpha$ 
solutions of \eqref{eq:eq_gamma}, $\gamma$ being the largest. 
Note that in this form we can also compare with Eq. (11) in \cite{Gueneau2025},
(there in the limit $N=+\infty$ first), the analogy between the two problems 
being even more striking.

\begin{figure}
   \centering
    \includegraphics[width=0.45\textwidth]{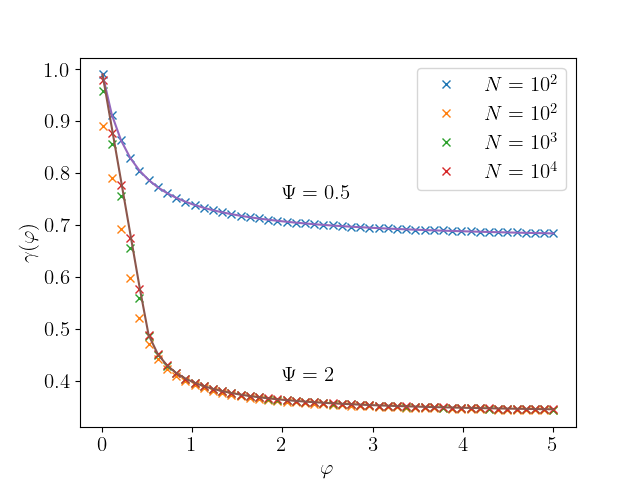}
    \caption{Plot of $\gamma(\varphi)$ obtained numerically (dots), compared to the analytical prediction (continuous lines),
    for $\sigma=0$ and rigid $m_i$ given by the quantiles of  $\rho(m)=\psi (1-m)^{\psi-1} \theta(0<m<1)$,
    for $\psi=1/2$ (no localised phase) and $\psi=2$.}
    \label{fig:gammamirigid}
\end{figure}


 \section{Numerical simulations}
 
 We solve numerically
a discrete time version of the equation of motion \eqref{eq:ODE_growth} in the variable $v_i= \log x_i$,
which reads
\begin{equation} 
\frac{{\rm d}v_i}{{\rm d}t} = (m_i - \varphi - \frac{\sigma^2}{2}) + \frac{\varphi}{N} \sum_{j=1}^N e^{v_j(t)-v_i(t)}+ \xi_i(t)
\end{equation} 
We study the stationary regime in the large time limit. 

In the case without noise, $\sigma=0$ and for the model 
$\rho(m)=\psi (1-m)^{\psi-1} \theta(0<m<1)$ 
with $m_>=1$,
our prediction for the growth rate is (i) in the delocalised phase for $\varphi>\varphi_c=\max(0,1 - \frac{1}{\psi})$,
$\gamma$ is solution of 
\begin{equation} \label{eq:predDelocalpsi}
\frac{\varphi \, _2F_1\left(1, 1; \psi + 1; \frac{1}{\gamma + \varphi}\right)}{\gamma + \varphi} = 1
\end{equation}
while in the localised phase, for $\varphi<\varphi_c$, $\gamma=1-\varphi$. This prediction
is checked in Fig. \ref{fig:gammamirigid}. Here the $m_i$ have been chosen to be rigid given by the quantiles of their distribution.

In Fig. \ref{fig:gammam} (Left) we check the prediction for the largest occupation ratio, $p_1=x_1/\sum_i x_i$,
given in \eqref{p1phi}. The value of $\varphi_c$ is consistent with finite size
corrections of $m_1$. 

\begin{figure}
   \centering
    \includegraphics[width=0.45\textwidth]{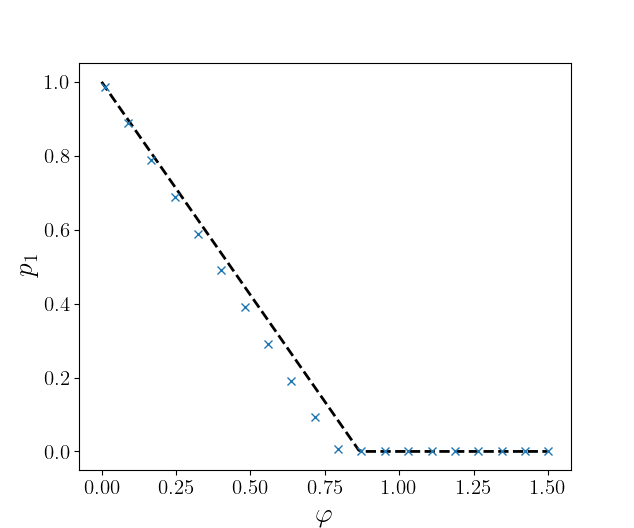}
    \includegraphics[width=0.5\textwidth]{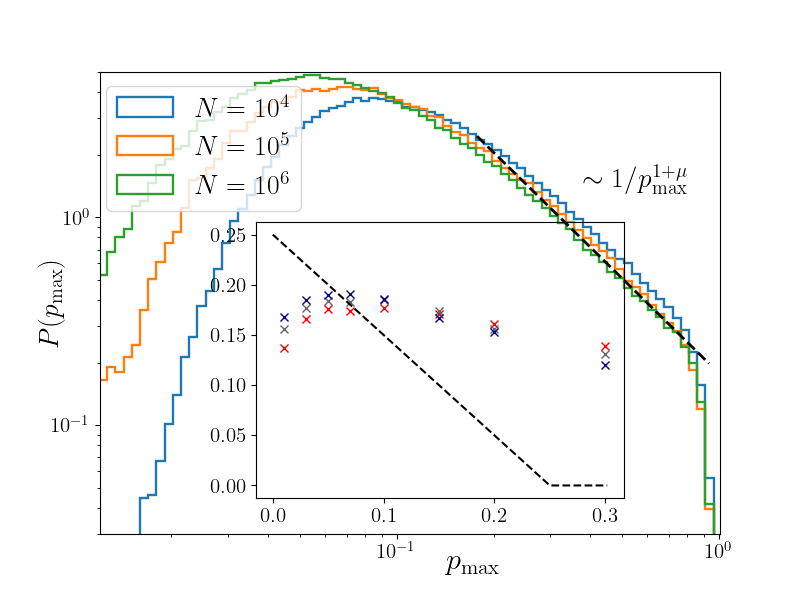}
    \caption{{\bf Left:} Plot of $p_1$ in the absence of noise $\sigma = 0$ obtained numerically (dots)
    for $N=10^4$ with $m_i$ as rigid variables set by the quantiles of a Gaussian distribution with $\Sigma_0 = 1$. 
    It is compared with our prediction, i.e. $p_1 =1 - \varphi/\varphi_c$ in the localised phase $\varphi<\varphi_c=m_1=\max_i m_i$,
    while it goes as $p_1 \sim 1/N $ in the delocalised phase. For $N=10^4$ one has
    $m_1 \approx 0.87$, {which slowly converges to $1$ as $N$ further increases}. {\bf Right:} Main: Plot of the probability distribution of $p_{\rm max}=\max_i p_i$ for various $N$ for $\varphi=0.15$, $\Sigma_0=1$ and $\sigma^2=1.44$. It corresponds to the partially localised phase. The distribution decays as a power-law with exponent predicted by the REM argument $\mu= 1 -\frac{\Sigma_0^2}{\sigma^4}$. Inset: Measured growth rate $\gamma$ for $\sigma^2=2$ as a function of $\varphi$ for various $N$ (dots), compared to the asymptotic prediction $\gamma_{\rm p.l.}=\frac{\Sigma_0^2}{2 \sigma^2}-\varphi$ (dashed lines).}
    \label{fig:gammam}
\end{figure}

In Fig. \ref{fig:gammam} (Right), we test the existence of the partially localised phase. We find that the distribution of $p_{\rm max}$ decays as a power-law $\sim 1/p_{\rm max}^{1+\mu}$, where $\mu$ is accurately predicted by the REM argument $\mu= 1 -\frac{\Sigma_0^2}{\sigma^4}$. 
In the inset, we compare the numerically obtained values of $\gamma$ with our asymptotic prediction. While logarithmic corrections in $N$ makes a quantitative comparison difficult, the $N$-dependence we observe is consistent with our analytical results.

 A more in-depth numerical study of the phases will be presented in an upcoming publication.

\end{widetext}


\begin{thebibliography}{}
\bibitem{footnotewishart} {Another simple case is when $\rho(m)$ corresponds to a Wishart distribution of parameter $q$, in which case $R(z)=(1-qz)^{-1}$ and $m_>=(1+\sqrt{q})^2$. One then finds $\gamma=\varphi/(\varphi-q)$ and $\varphi_c = \sqrt{q}/(1+\sqrt{q})$.}
\bibitem{footnote1}
From the expansion \eqref{expansionfixedN}, noting that the next order correction
is only $O(1/(\varphi^3 (\log N)^2))$.
\bibitem{footnotem1random}
{{Note that for finite $N$, $m_1$, and consequently $\gamma$, exhibits fluctuations that vanish as $N \to \infty$. These fluctuations have been averaged out here.}}
\bibitem{footnoteGabaixIneq}
In this paper, we consider the case where the fluctuations $\xi_i(t)$ are delta-correlated in time. The case of persistent fluctuations has been considered in a different context in \cite{gabaixIneq}. In fact, another way to interpolate between static disorder ($m_i$) and white noise ($\xi_i(t)$) is to introduce a non-zero correlation time $\tau \propto \log N$. We will consider such an extension in a future work.
\bibitem{footnotefigpmax}
Note that we have considered here the $m_i$ rigid, given by the quantiles of a Gaussian distribution with $\Sigma_0=1$. The fluctuations shown are thus with respect to the noise $\sigma>0$.
\bibitem{footnotefluct}
Note that we have neglected here the fluctuations of $\bar x$. These, and their possible effect on the phase diagram, will be studied in an upcoming publication.
\bibitem{footnotecontI1}
If this condition is not satisfied, $I_1$ is dominated by the vicinity $v=1$. This contribution turns out to be irrelevant as it happens in the localised/partially localised phase, and $I_2 \gg I_1$ in this case.
\bibitem{footnoteconcl}
These properties will be detailed in an upcoming publication.
\end{thebibliography}
\end{document} 

